\documentclass[aps,prl,twocolumn]{revtex4}
\usepackage{graphicx,amsmath,amssymb,bm}

\begin{document}

\title{Theory of the skyrmion, meron, anti-skyrmion and anti-meron in chiral magnets} 
\author{Sandip Bera$^1$ and Sudhansu S. Mandal$^{1,2}$ }
\affiliation{$^1$Department of Physics, Indian Institute of Technology, Kharagpur 721302, India\\ 
$^2$Centre for Theoretical Studies, Indian Institute of Technology, Kharagpur 721302, India	
	}

\date{\today}

\begin{abstract}
	
We find closed-form solution of the Euler equation for a chiral magnet in terms of a skyrmion or a meron depending on the relative strengths of magnetic anisotropy and magnetic field. We show that the relevant length scales for these solutions  primarily depend on the strengths of Dzyaloshinskii-Moriya interaction through its ratios, respectively, with magnetic field and  magnetic anisotropy.  We thus unambiguously determine the parameter dependencies on the radius of the topological structures particularly of the skyrmions, showing an excellent agreement with experiments and first-principle studies. 
An anisotropic Dzyaloshinskii-Moriya interaction suitable for thin films made with $C_{nv}$ symmetric materials is found to stabilize anti-skyrmion and anti-meron, which are prototypical for $D_{2d}$ symmetric systems, depending on the degree of anisotropy. Based on these solutions, we obtain phase diagram by comparing the energies of various collinear and non-collinear competing phases.

\end{abstract}

\maketitle


 The chiral Dzyaloshinskii-Moriya interaction (DMI) \cite{D57,M60} for broken inversion symmetric systems is one of the most important mechanisms including frustrated exchange interactions and long-ranged dipolar interaction for producing one-dimensional modulation in magnetization known as spin-spiral \cite{Bogdanov89,Bogdanov06,Meckler09,Ezawa,Leonov16,Zhang17,Okubo12} in ferromagnetic systems. An application of magnetic field in such a system stabilizes \cite{Bogdanov94,Leonov16} skyrmions (Sks) having topologically protected quasiparticle-like spin structure \cite{Muhlbauer09,Tokura10a,Heinze11,Tokura12,Romming13,Tanigaki15} in a ferromagnetic background. N\'eel and Bloch type Sks are generally realized  respectively in $C_{nv}$ and $D_n$ symmetric \cite{Bogdanov89,Bogdanov94} bulk and thin-film materials for wide range of  magnetic fields and temperatures \cite{Muhlbauer09,Tokura10a,Heinze11,Tokura12,Romming13,Tanigaki15,Kezsmarki15,Fujima17,Tokura10,Chacon,Das19}.

Recent observations \cite{Nayak17,Nayak19} of anti-skyrmions (ASks) in Heusler alloys with $D_{2d}$ crystal symmetry have raised an issue about the microscopic environment which will stabilize a Sk or an ASk. While a Sk has either N\'eel or Bloch type of orientation of magnetization vector governed by the respective transverse and longitudinal DMI, an ASk displays a combination of both. It is thus tempting to think that an anti-skyrmion may be produced in a crystal whose symmetry gives rise to both types of DMI. Numerical simulations, on the contrary, \cite{Koshibae14,Koshibae16,Camosi18} indicate that the ASks do  stabilize only in the presence of dipolar interaction. A micromagnetic study \cite{Hoffmann17} suggests that Sks and ASks can, however, coexist and this coexistence is predicted by electronic structure calculation at interfaces due to anisotropic DMI. These ASks even take part in current-induced motion \cite{Huang17}.

Hoffmann et al \cite{Hoffmann17} have recently observed ASks in $C_{2v}$ symmetric systems grown on semiconductor or heavy-metal substrates, while $C_{nv}$ symmetric systems are known to stabilize Sks only \cite{Bogdanov89,Leonov16}.  This motivates us to study a system of thin film chiral magnet that may be fabricated with $C_{nv}$ symmetric crystals with an anisotropic DMI in a continuum model in search of ASk solution. Camosi et al \cite{Camosi17} have recently reported that the epitaxially grown thin Co films on W(110) brings anisotropy in DMI along two orthogonal growth directions of a $C_{2v}$ symmetric bulk system. Although this reported anisotropy does not correspond to two opposite signs along two orthogonal directions, a micromagnetic simulation seems to suggest anisotropy in thin films not only in magnitude but also in sign \cite{Huang17}. We introduce a model  DMI with such an anisotropy.

Moreover, recent observation of another topological spin structure, {\it viz}, meron \cite{Nych17,Tokura18} have further raised the theoretical issue on the parameter regimes on which all these different kinds of topological structures emerge. Further, definite parameter dependencies on the radius \cite{Romming15,Malottki,Thiaville,Wang} and appropriate length scale \cite{Leonov16} of a SK are not yet settled.
Our focus is thus  solving basic Euler equation for angular variables representing magnetization with isotropic DMI for Sks and merons and then study the consequences of its anisotropy followed by the determination of phase diagram by comparing energies of different possible solutions for thermodynamically stable magnetic structures.

In this letter, we solve the Euler equation in a continuum model \cite{Bogdanov89,Leonov16} 
with ferromagnetic exchange coupling, $J$, DMI strength, $D$, strength of magnetic anisotropy, $A$, and net Zeeman energy due to magnetic field, $H$.
For moderate to high $HJ/D^2$ and $\gamma=2A/H < 1$, we find that the relevant length scale of the corresponding skyrmion solution is $r_s = D/H$, contrasting the belief \cite{NT} of the relevant length scale $r_d = J/D$. This enables us to determine the magnetic field and anisotropy dependencies of the radius of a skyrmion and find that it is in excellent agreement with experiments \cite{Romming15,Vousden16} and first-principle studies \cite{Malottki}.   
The meron solution at zero magnetic field is obtained for $A>0$ (easy-plane anisotropy) by minimizing energy and the relevant length scale is found to be $D/A$. We show the formation of meron lattice and argue how a symmetric Sk is evolved from a meron via an asymmetric skyrmion, explaining a recent experiment as well as simulation results\cite{Tokura18}. Further, our model with an anisotropic DMI is shown to stabilize ASks and antimerons in $C_{nv}$ symmetric systems, as evident in recent realization \cite{Hoffmann17} of ASks in $C_{2v}$ symmetric systems. We finally determine phase diagram for $\gamma <1$ by comparing energies of the skyrmion solution with other collinear and non-collinear competing phases.


We begin with considering a two-dimensional chiral magnet having energy $E = \int d^2r ({\cal E}_{{\rm EX}} +{\cal E}^{\pm}_{{\rm DM}}+{\cal E}_{{\rm AH}})$ with respect to an overall ferromagnet orienting along perpendicular to the plane of the system, described by exchange energy density ${\cal E}_{{\rm EX}} = \frac{J}{2} \left( \nabla \hat{m} \right)^2$, DMI  energy density ${\cal E}^{\pm}_{{\rm DM}} = -D \left( L_{xz}^{(x)} \pm L_{yz}^{(y)} \right) $. Here $L_{ij}^{(k)} = \hat{m}_i \partial_{x_k} \hat{m}_j - \hat{m}_j \partial_{x_k} \hat{m}_i$, $\hat{m}$ is unit magnetization vector, $\pm$ signs, respectively, refer to the systems with $C_{nv}$ and $D_{2d}$ symmetries when the Dzyaloshinskii-Moriya vector is transverse \cite{fnote1} to the lattice-bond. (While the former supports SKs the later is suitable for stabilizing ASks.) The energy density for magnetic anisotropy and applied magnetic field along $\hat{z}$ direction given by ${\cal E}_{{\rm AH}} = -A\, (1-\hat{m}_z^2) + H \,(1-\hat{m}_z)$, where $A>0\, (<0)$ refers to easy-plane (easy-axis) anisotropy. In spherical polar representation,
$$\hat{m}({\bm{r}}) =\left[ \cos \Phi (\bm{r}) \sin \Theta (\bm{r}),\, \sin \Phi (\bm{r}) \sin \Theta (\bm{r}),\, \cos \Theta (\bm{r}) \right]$$
with $\bm{r} = (r\cos\phi, r\sin\phi)$ in polar coordinate system.

  A topological structure defined by its topological quantum number $N_{{\rm sk}} = \frac{1}{4\pi} \int d^2r \,\hat{m} \cdot \left( \partial_x \hat{m} \times \partial_y \hat{m} \right) \equiv \pm N_0$, 
 where positive sign  refers to a Sk or meron and negative sign refers to an ASk or anti-meron.
 The solutions \cite{Leonov16} of a Sk/meron and an ASk/anti-meron correspond to $\Theta (\bm{r}) = \Theta (r)$ and, respectively, $\Phi (\bm{r}) = \pm \phi + \eta$. 
 Here $\eta$ determines a constant extra planar rotation of magnetic moment at all points; $\eta = 0\,(\pi/2) $  for Ne\'el(Bloch) type topological structures. 
 Here $N_0$ represents the winding number \cite{NT}:  its positive (negative) sign determines inward (outward) spin orientation with respect the origin, corresponding to negative (positive) sign of $D$, 
 and its magnitude is $1$ for Sks and ASks, and $1/2$ for merons and anti-merons.   
 The boundary condition, $\hat{m} = (0,0,+1)$ for $r\to \infty$, {\it i.e.}, $\Theta (r\to \infty) = 0$ and $\hat{m} = (0,0,-1)$ at $r=0$, {\it i.e.,} $\Theta (r=0) = \pi$ is for both SK and aSK.  Meron and anti-meron correspond to the boundary condition $\Theta (r=\infty) = \pi/2$ and $\Theta (r=0) = \pi\, (0)$ for inward (outward) helicity.
 
\begin{figure}
	\centering	\includegraphics[width=8.5cm]{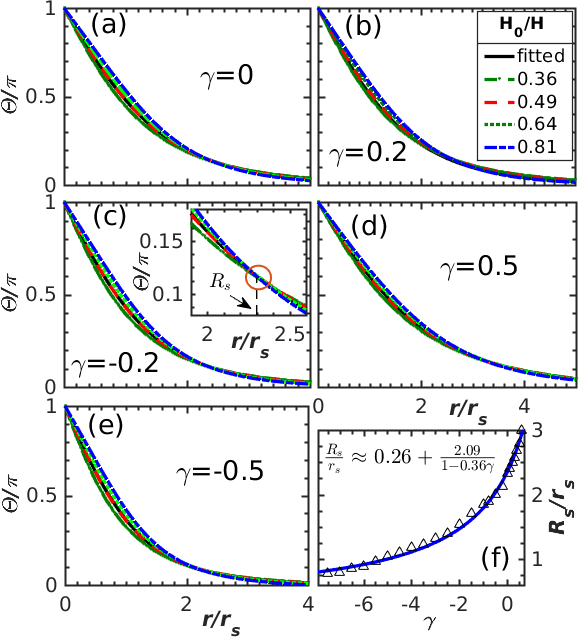}
	\caption{(color online) Skyrmion or anti-Skyrmion solution: Numerical solution of the Euler equation (\ref{Euler}), {\it i.e.}, $\Theta (r)$ {\it vs.} $r/r_s$ for different values of $\gamma$ in the panels (a)--(e) for same set of $H_0/H$, {\it viz}, 0.36, 0.49, 0.64 and 0.81. Solid lines represent the best fit solution with the form given by Eq.~(\ref{S0}). Inset in the panel (c) is a representative of the panels (a)--(e) for showing the crossing of the curves for different values of $H_0/H$. Also see supplemental material \cite{Suppli} for wider range of parameters. This crossing point is identified as the radius, $R_s$,  of a skyrmion.  The dependence of $R_s$ on $\gamma$ along with its approximate fitted form is shown in the panel (f).   }
	\label{Fig1}
\end{figure}

 No matter, be it $C_{nv}$, $D_{2d}$ or $D_n$ systems, the Euler equation for $\Theta (r)$ is identical \cite{Suppli}. 
 By introducing a length scale $r_s = D /H$ and rescaling $r \to r_s \rho$, we obtain \cite{Suppli} the Euler equation
 \begin{eqnarray}
 &&\frac{d^2\Theta}{d\rho^2} + \frac{1}{\rho}\frac{d\Theta}{d\rho}- \frac{\sin \Theta \cos \Theta}{\rho^2}   \nonumber \\
 && =  \frac{H_0}{H} \left( -\frac{2}{\rho}\sin^2\Theta 
 +\sin\Theta - \gamma \sin\Theta \cos\Theta \right)   
 \label{Euler}
 \end{eqnarray}
  where $H_0 = D^2/J$ and $\gamma = 2A/H$.
 The numerical solutions of the Eq.~(\ref{Euler}) with the boundary conditions $\Theta (0) = \pi$ and $\Theta (\infty) = 0$ for different values of $H_0/H$ and $\gamma$ are shown in Fig.~1.  The length scale $r_{s}$ which is independent of exchange energy $J$ defines the relevant length scale as for a fixed value of $\gamma$, the deviation of the curves of $\Theta (r)$ for different values of $H_0/H$ are almost negligible; the complete solution of Eq.(\ref{Euler}) is thus best approximated by    
 \begin{equation}
 	\Theta (r) = 4 \arctan \left(\exp \left(-\beta(\gamma)r/r_s \right) \right)
 	\label{S0}
 \end{equation}
 with $\beta(\gamma) \approx 0.91 - 0.55 \gamma$. 
 We note that all the curves for a fixed $\gamma$ cross (see inset of Fig.~\ref{Fig1}(c)) at a particular $r$ and we identify that to be the radius, $R_s$,  of a Sk.
 We find its dependency on $\gamma$ as $R_s = r_s w(\gamma)$ with  $w(\gamma) \approx 0.26 +\frac{2.09}{1-0.36\gamma}$. Therefore, the magnetic field dependence of the radius of a Sk may be parametrized as 
 \begin{equation}
 R_s = \frac{C_1}{H} + \frac{C_2}{H-C_3}
 \label{Size}
 \end{equation}
  where the coefficients $C_1$ and $C_2$ are proportional to $\vert D \vert$ and $C_3$ is proportional to $A$. We note that for a fixed $H$, radius of a Sk increases with positive $A$, in agreement with an experiment \cite{Vousden16}. However, an increase of easy-axis anisotropy will reduce the size of an Sk.
  Figure 2 shows that the skyrmion radius obtained in an experiment \cite{Romming15} and first-principle studies \cite{Malottki} obey the relation (\ref{Size}) very well and the sign of the corresponding fitted $C_3$ are consistent with the sign of the reported $A$. 
   For the systems with positive $A$, lower bound of the magnetic field needed for producing a Sk is $H_{{\rm lb}} = 2A$ and thereafter the radius monotonically decreases with increasing $H$.

  \begin{figure}
  	\centering	\includegraphics[width=8.0cm]{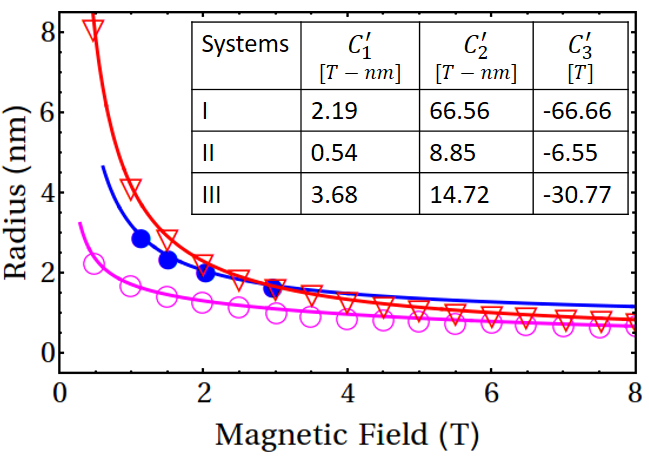}
  	\caption{(color online) Variation of radius of a Sk with applied magnetic field: Experimental data (solid circles) from Ref.~\onlinecite{Romming15}, first-principle calculations from Ref.~\onlinecite{Malottki} for hcp lattice (open circle) and fcc lattice (open inverted triangles), denoted respectively as I, II and III. Solid lines are the fitted curves of these data with the functional form in Eq.~\ref{Size}. Inset: coefficients obtained by fitting are tabulated where we have put primes to distinguish from non-primes in Eq.(\ref{Size}) in view of dimensions.  }
  	\label{Fig2}
  \end{figure} 
  
  Figure \ref{Fig3}(a) shows phase diagram in $A$--$H$ space with $\gamma <1$. The phase boundary between skyrmion and the polarized ferromagnet is determined by comparing energy of a Sk,
  \begin{eqnarray}
  & & E_{{\rm sk}} = 2\pi \int_0^\infty dr\, r \left[  \frac{J}{2} \left(  \left( \frac{d\Theta}{dr} \right)^2 +\frac{\sin^2\Theta}{r^2} \right) -A \sin^2\Theta  \right. \nonumber \\
  & &  \hspace{0.3cm} \left.  +D \left(   \frac{d\Theta}{dr}   +\frac{\sin\Theta \cos \Theta}{r}  \right)+ H (1-\cos\Theta)     
  \right]  \label{Energy_sk} 
  \end{eqnarray}
  with the energy of the ferromagnet. Similarly by determining energy of a spin-spiral following Ref.~\onlinecite{Bogdanov89} in comparison to the ferromagnet, we obtain the phase-boundary between the spin-spiral and ferromagnet. We draw phase boundary between spin-spiral and skyrmion phases by considering maximum possible phase-space for spin-spiral structure. The phase diagram for $\gamma <1$ here is consistent with previously reported phase diagrams obtained by variational and other simulations \cite{Banerjee14,Rowland16}. In agreement with an experiment \cite{Herve18}, both spin-spiral and skyrmions are accessible at zero anisotropy.

  \begin{figure}
  	\centering \includegraphics[width=8.5cm]{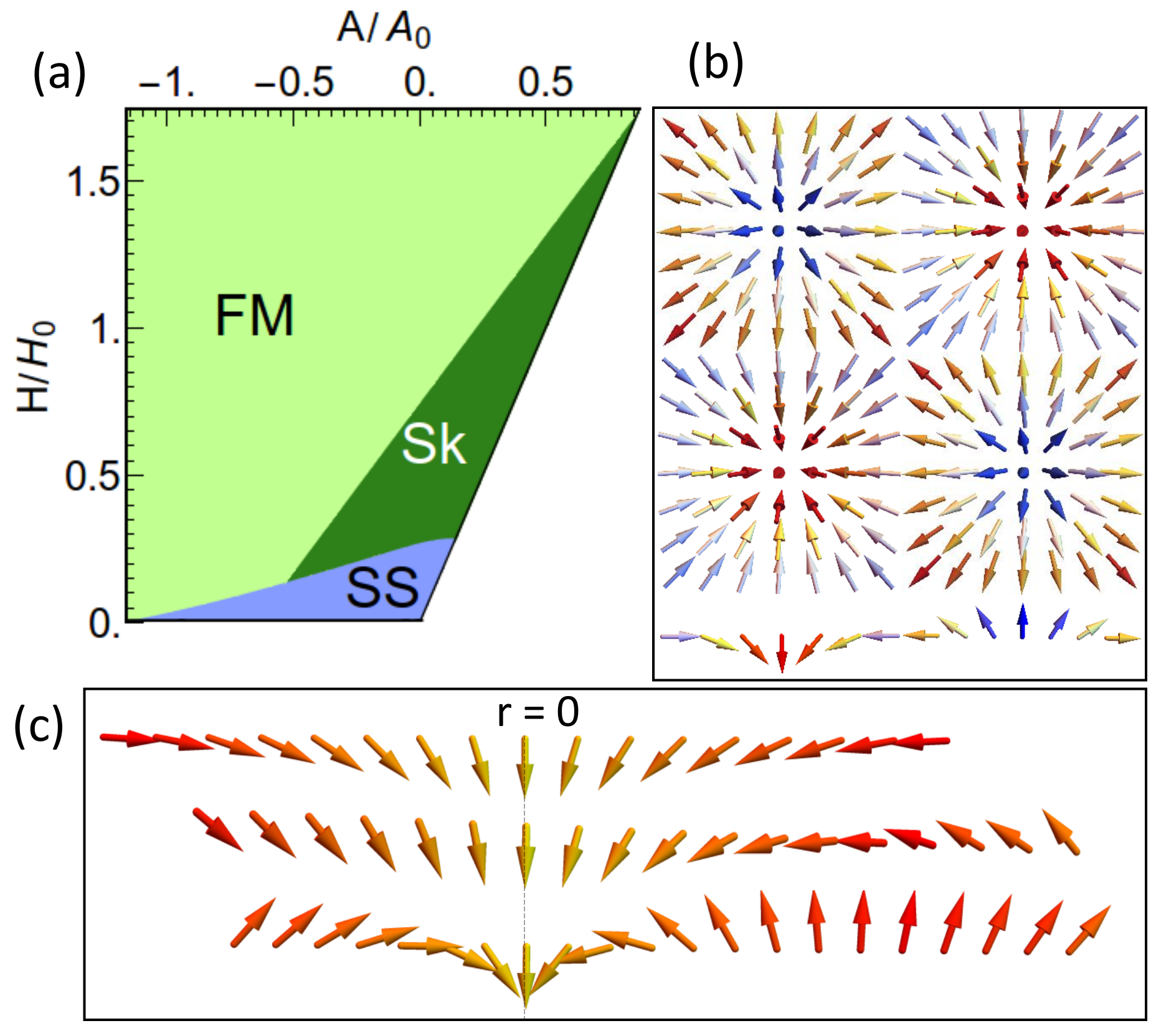}
  	\caption{(color online) (a) Phase Diagram in $A/A_0$--$H/H_0$ space, where $A_0 = H_0 = D^2/J$. Three distinct phases:  Spin-spiral (SS), isolated skyrmion (SK), and  polarized ferromagnet (FM) whose  magnetization direction is along the applied $H$. The right boundary corresponds to $\gamma = 1$. (b) Two possible degenerate structures of merons (up or down spin at the center) for $H=0$. From top, two and one-dimensional spin-structures of merons. Any finite $H$ stabilizes meron with down-spin at the center only. (c) From top, depiction of a symmetric meron ($H=0$), an asymmetric meron ($H>0$, $\gamma >1$), and an asymmetric skyrmion. Spin-down at $r=0$ for all these structures. Here the schematics (b,c) of merons and skyrmion are considered for $D<0$.
  	}
  	\label{Fig3}
  \end{figure}

  For a sufficiently high easy-plane anisotropy $(A>0)$ and $H=0$, all the spins will align in the plane (planar ferromagnet). This indicates a boundary condition $\Theta (\infty) = \pi/2$ which together with another boundary condition $\Theta (0)= 0 $ or $\pi$ will provide a solution of meron when $A$ is moderate. Taking cue of the skyrmion solution, we assume the solutions of meron \cite{Suppli} to be
  \begin{equation}
  \Theta (r) = \pm \frac{\pi}{2} + 2 \arctan (\exp(-\zeta r/r_a))
  \end{equation}
  where $r_a =  D  /A$ is the characteristic length scale, positive (negative) sign corresponds to spin down (up) at the center of the meron, and the parameter $\zeta $ to be determined by minimizing its energy
  \begin{eqnarray}
  E_{{\rm meron}} &=& 2\pi\int_0^\infty r \, dr  \left[ \frac{J}{2} \left( \left( \frac{d\Theta}{dr}\right)^2 +\frac{\sin^2\Theta}{r^2}  \right)   \right. \nonumber \\
  && + A \cos^2\Theta
  +\left. D \left( \frac{d\Theta}{dr} + \frac{\sin \Theta\cos \Theta}{r} \right) \right] \, .
  \label{Ener_meron}
  \end{eqnarray}
 We find \cite{Suppli} $\zeta  = 2\ln(2)/(1+2G)\approx 0.49$, where $G$ is Catalan's constant. These solutions of $\Theta(r)$ are degenerate and hence they occur simultaneously and appear as neighbors to match the background of planar ferromagnet and form a meron-lattice, as shown in Fig.~3(b). However, with the increase of $H$, only one-kind of meron (spin-down at its core) survive as the other will have higher energy, because the background of spin alignment will have nonzero out of plane (up) component. For further increase of $H$, this meron gradually converts into a skyrmion as it helps to orient more spin with finite up component.  This is in reminiscent of the recently observed merons by Yu {\it et al.} \cite{Tokura18}. We estimate the upper-bound of $A$ for forming a meron as $A_{{\rm ub}} \approx 2.3 A_0$ by comparing the energy of a meron and the planar ferromagnet.

  In the presence of $H$ with  $A>>0$ such that $\gamma >1$, a tilted ferromagnet will be formed with finite amount of spin-projection along the direction of $H$ making a tilting angle $\arccos (1/\gamma)$ with the plane. With such a tilted ferromagnet in the background, locally formed merons with down-spin at its core will be asymmetric as shown schematically in Fig.~3(c) when $A<A_{{\rm ub}}$. If we look along a particular direction, a meron's spin alignment at one boundary will be along the tilting angle $\arccos (1/\gamma)$ and at the other boundary it will differ by an angle $\pi$. This makes the meron asymmetric. We note that actual $A_{{\rm ub}}$ may be lower than estimated here because of the predicted possibility of forming cone-like structure in the intermediate regime. The cone structure \cite{Rowland16,Banerjee14} and the tilted-ferromagnet are indistinguishable in our analysis because both these structures correspond to same $\Theta$.  With further increase of $H$, some more spins will tend to align more than $\arccos (1/\gamma)$ forming an asymmetric skyrmion (Fig.~3(c)), corroborated with the recent numerical simulation result \cite{Leonov17}. Upon further increase of $H$, right(left) side of the Sk becomes shorter(longer) and evolve into a symmetric Sk at $\gamma =1$ as we enter into the Sk phase of the phase-diagram (Fig.3(a)).

 In search of ASk and anti-meron in thin films made of $C_{nv}$ symmetric sytems \cite{fnote2}, we introduce an anisotropic DMI given by
  \begin{equation}
  {\cal E}_{{\rm DM}}^+ = -D (1+\lambda \cos 2\phi) \left( L_{xz}^{(x)} + L_{yz}^{(y)} \right) \, .
  \end{equation}
   Here $\lambda$ denotes the degree of anisotropy with $\lambda =0$ representing the symmetric DMI present in the bulk $C_{nv}$ symmetric crystals. The energy of an ASk is then found to be
\begin{eqnarray}
& & E_{{\rm ask}} = 2\pi \int_0^\infty dr\, r \left[  \frac{J}{2} \left(  \left( \frac{d\Theta}{dr} \right)^2 +\frac{\sin^2\Theta}{r^2} \right) -A \sin^2\Theta  \right. \nonumber \\
& &  \hspace{0.3cm} \left. 
 +\frac{\lambda}{2} D \left(   \frac{d\Theta}{dr}   -\frac{\sin\Theta \cos \Theta}{r}  \right)  + H (1-\cos\Theta)        
\right]  \label{Energy_ask}
\end{eqnarray}
with $\Theta (r)$ given in Eq.~(\ref{S0}).
 Inset of Fig.~\ref{Fig4}(a) shows the variation of $E_{{\rm ask}}$ with $\lambda$ for $A=0$ and we find that $E_{{\rm ask}} < E_{{\rm sk}}$ above a critical value $\lambda_c \approx 1.4$ and hence the anisotropy in DMI stabilizes an ASk. A phase diagram has been presented in Fig.~\ref{Fig4}(a). Ferromagnet to ASk transition is also possible for $\lambda >\lambda_c$, and the corresponding critical value increases with $H$. However, ASks are not possible for lower $H/H_0$ where spin-spiral phase remain unaltered for any $\lambda$. 
 Figure \ref{Fig4}(b) shows minimum values of $\lambda$ above which the full phase-space of Sks and partial-phase of ferromagnets shown in Fig.~\ref{Fig3}(a) can stabilize ASks.
 The outer boundary in the ferromagnetic region is obtained with the criterion that the ratio of the diameter of an ASk and the spin-spiral wavelength is not less than 0.4.

 The energy of an anti-meron in presence of anisotropic DMI,
 \begin{eqnarray}
 & &E_{{\rm anti-meron}} = 2\pi\int_0^\infty r \, dr  \left[ \frac{J}{2} \left( \left( \frac{d\Theta}{dr}\right)^2 +\frac{\sin^2\Theta}{r^2}  \right)   \right. \nonumber \\
 && + A \cos^2\Theta
 +\left. \frac{\lambda}{2} D \left( \frac{d\Theta}{dr} - \frac{\sin \Theta\cos \Theta}{r} \right) \right] \, .
 \label{Ener_antimeron}
 \end{eqnarray}
 becomes less than $E_{{\rm meron}}$ for $6.8 \lesssim \lambda$. Producing anti-meron by anisotropic DMI is less probable than producing ASk because the former requires much higher degree of anisotropy, which is almost in the verge of the limit of a $D_{2d}$ system.

  We here have shown that the anisotropic DMI in thin films with $C_{nv}$ symmetric materials can host anti-skyrmions for wide range of phase-space of $A$ and $H$, in comparison to hosting skyrmions. However, we do not find any regime of the coexistence of Sks and ASks, in contrary to the numerical simulation \cite{Hoffmann17}. Although dipolar interaction is also a suitable mechanism \cite{Koshibae14,Koshibae16,Camosi18}for stabilizing ASks, the anisotropic DMI is solely responsible, to the best of our knowledge, for small-size ASks in $C_{nv}$ symmetric systems. The dipolar interaction here may play a role in reducing \cite{Lobanov16} the effect of magnetic anisotropy.
  The physics of Sk/ASk and meron/anti-meron discussed here will reverse for systems with $D_{2d}$ symmetries. Although the structure of an anti-skyrmion is a combination of the  structures of N\'eel and Bloch type Sks which are prototypical, respectively, of DMI with Dzyaloshinskii-Moriya vector orthogonal to the neighboring bond and along the bond, their combinations do not produce ASks. However, a pure $D_n$ symmetric system will stabilize Bloch type merons and SKs, and the corresponding anti-merons and ASks may also be produced through anisotropic DMI.

  S.S.M. is supported by SRIC, IIT Kharagpur through the project code EFH.

 	\begin{figure}
 		\centering	\includegraphics[width=8.5cm]{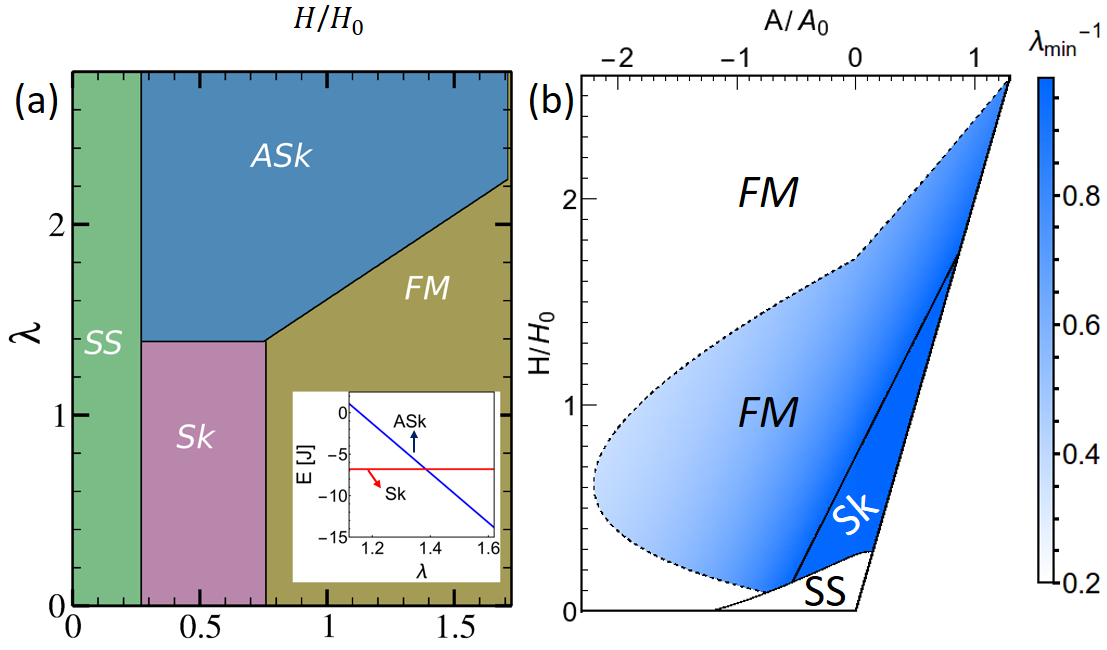}
 		\caption{(color online) (a) Phase diagram in $H/H_0$--$\lambda$ plane when $A=0$. Four distinct phases are obtained by comparing their energies. While the energies of a skyrmion and an anti-skyrmion have been calculated using the expressions (\ref{Energy_sk}) and (\ref{Energy_ask}), respectively,  the energy of spin-spiral phase has been obtained following the procedure reported in Ref.~\onlinecite{Bogdanov89}.
 	The inset shows comparison of the energies of a Sk and an ASk with respect to the energy of the ferromagnetic ground state in the back-ground as a function of anisotropy parameter $\lambda$ of DMI. The crossing point of two lines determines the boundary between skyrmionic and anti-skyrmionic phases.  (b) The color map indicates minimum value of $\lambda$ above which an ASk is stabilized over a Sk or a ferromagnet in the parameter space shown in Fig.~\ref{Fig3}(a).  }
 		\label{Fig4}
 	\end{figure}





\end{thebibliography}


\begin{thebibliography}{99}
	
	
\bibitem{D57} I. E. Dzyaloshinskii,  Zh. Eksp. Theo. Fiz. {\bf 32}, 1547 (1957)[Sov. Phys. JETP {\bf 5}, 1259 (1957)].
	
		
\bibitem{M60} T. Moriya,  Phys. Rev. {\bf 120}, 91 (1960).
	

\bibitem{Bogdanov89} A. N. Bogdanov and D. A. Yabionskii,  Sov. Phys. JETP {\bf 68}, 101 (1989).


\bibitem{Bogdanov06} U. K. R\"o{\ss}ler, A. N. Bogdanov, and D. Pfleiderer,  Nature {\bf 442}, 797 (2006).



\bibitem{Meckler09} S. Meckler, N. Mikuszeit, A. Pre{\ss}ler, E. Y. Vedmedenko, O. Pietzsch, and R. Wiesendanger, Phys. Rev. Lett. {\bf 103}, 157201 (2009).	

\bibitem{Ezawa} M. Ezawa, Phys. Rev. Lett. {\bf 105}, 197202 (2010).	


\bibitem{Leonov16} A. O. Leonov, T. L. Monchesky, N. Romming, A. Kubetzka, A. N. Bogdanov, and R. Wiesendanger,  New J. Physics {\bf 18}, 065003 (2016).


\bibitem{Zhang17} X. Zhang, J. Xia, Y. Zhou, X. Liu, H. Zhang and M. Ezawa, Nature Comm. {\bf 8}, 1717 (2017)

\bibitem{Okubo12} T. Okubo, S. Chung, and H. Kawamura, Phys. Rev. Lett. {\bf 108}, 017206 (2012).

\bibitem{Bogdanov94} A. Bogdanov and A. Hubert,  J. Mag. Mag. Mat. {\bf 138}, 255 (1994).	

\bibitem{Muhlbauer09}S. M\"uhlbauer, B. Binz, F. Jonietz, C. Pfleiderer, A. Rosch, A. Neubauer, R. Georgii, and P. B\"oni, Science {\bf 323}, 915 (2009).


\bibitem{Tokura10a}  X. Z. Yu, N. Kanazawa, Y. Onose, K. Kimoto, W. Z. Zhang, S. Ishiwata, Y Matsui and Y. Tokura, Nature Materials. {\bf 10}, 106 (2011).


	
\bibitem{Heinze11} S. Heinze, K. von Bergmann, M. Menzel, J. Brede, A. Kubetzka, R. Wiesendanger, G. Bihlmayer, and S. Bl\"ugel, Nature Phys. {\bf 7}, 713 (2011).
	
	
\bibitem{Tokura12} A. Tonomura, X. Yu, K. Yanagisawa, T. Matsuda, Y. Onose, N. Kanazawa, H. S. Park, and Y. Tokura, Nano Lett. {\bf 12}, 1673 (2012).

\bibitem{Romming13} N. Romming, C. Hanneken, M. Menzel, J. E. Bickel, B. Wolter, K. von Bergmann, A. Kubetzka, and R. Weisendamger, Science {\bf 341} 636 (2013).

\bibitem{Tanigaki15} T. Tanigaki, K. Shibata, N. Kanazawa, X. Yu, Y. Onose, H. S. Park, D. Shindo, and Y. Tokura, Nano Lett. {\bf 15}, 5438 (2015).


\bibitem{Kezsmarki15} I. K\'ezsm\'arki, S. Bord\'acs, P. Milde, E. Neuber, L. M. Eng, J. S. White, H. M. Ronnow, C. D. Dewhurst, M. Mochizuki, K. Yanai, H. Nakamura, D. Ehlers, V. Tsurkan and A. Loidl , Nature Materials. {\bf 14}, 1116 (2015).	


\bibitem{Fujima17} Y. Fujima, N. Abe, Y. Tokunaga, and T. Arima, Phys. Rev. B {\bf 95}, 180410(R) (2017).


\bibitem{Tokura10} X. Z. Yu, Y. Onose, N. Kanazawa, J. H. Park, J. H. Han, Y. Matsui, N. Nagaosa, and Y. Tokura, Nature {\bf 465}, 901 (2010).	
	
	 	
	
\bibitem{Chacon} A. Chacon, L. Heinen, M. Halder, A. Bauer, W. Simeth, S. M\"uhlbauer, H. Berger, M. Garst, A. Rosch and C. Pfleiderer, Nature Phys. {\bf 14}, 936 (2018). 
	
\bibitem{Das19} S. Das {\it et al.}, Nature {\bf 568}, 368 (2019).	
	
	
	
\bibitem{Nayak17} A. K. Nayak, V. Kumar, T. Ma, P. Werner, E. Pippel, R. Sahoo, F. Damay, U. K. R\"o{\ss}ler, C. Felser, S. S. P. Parkin,  Nature {\bf 548}, 561 (2017).

\bibitem{Nayak19} S. Jamaluddin, S. K. Manna, B. Giri, P. V. P. Madduri, S. S. P. Parkin, and A. K. Nayak, Adv. Funct. Mater. {\bf 0}, 1901776 (2019).


\bibitem{Koshibae14} W. Koshibae and N. Nagaosa,  Nature Comm. {\bf 5}, 5148 (2014). 


\bibitem{Koshibae16} W. Koshibae and N. Nagaosa,  Nature Comm. {\bf 7}, 10542 (2016).

\bibitem{Camosi18} L. Camosi, N. Rougemaille, O. Fruchart, J. Vogel, and S. Rohart, Phys. Rev. B {\bf 97}, 134404 (2018). 

 
\bibitem{Hoffmann17} M. Hoffmann, B. Zimmermann, G. P. M\"uler, D. Sch\"urhoff, N. S. Kiselev, C. Melcher, and S. Bl\"ugel,  Nature Comm. {\bf 8}, 308 (2017).

\bibitem{Huang17} S. Huang, C. Zhou, G. Chen, H. Shen, A. K. Schmid, K. Liu, and Y. Wu,  Phys. Rev. B {\bf 96}, 144412 (2017).


\bibitem{Camosi17} L. Camosi, S. Rohart, O. Fruchart, S. Pizzini, M. Belmeguenai, Y. Roussinn\'e, A. Stashkevich, S. M. Cherif, L. Ranno, M. Santis, J. Vogel,  Phys. Rev. B {\bf 95}	, 214422 (2017).
	
\bibitem{Nych17} A. Nych, Jun-ichi Fukuda, U. Ognysta, S. Zumer and I. Musevic,  Nature Phys. {\bf 13}, 1215 (2017).	



\bibitem{Tokura18} X. Z. Yu, W. Koshibae, Y. Tokunaga, K. Shibata, Y. Taguchi, N. Nagaosa and Y. Tokura, Nature {\bf 564}, 95 (2018).

\bibitem{Romming15} N. Romming, A. Kubetzka, C. Hanneken, K. von Bergmann, and R. Wiesendanger,  Phys. Rev. Lett. {\bf 114}, 177203 (2015).


\bibitem{Malottki} S. von Malottki, B. Dup\'e, P. F. Bessarab, A. Delin and S. Heinze, Sci. Rep. {\bf 7}, 12299 (2017)


\bibitem{Thiaville} S. Rohart and A. Thiaville, Phys. Rev. B {\bf 88}, 184422 (2013).


\bibitem{Wang} X. S. Wang, H. Y. Yuan and X. R. Wang, Comm. Phys. {\bf 1}, 31 (2018).	

%
\bibitem{NT} N. Nagaosa and Y. Tokura, Nat. Nanotechnol. {\bf 8}, 899 (2013).

\bibitem{Vousden16} M. Vousden, M. Albert, M. Beg, Marc-Antonio Bisotti, R. Carey, D. Chernyshenko, D. Cortes-Ortuno, W. Wang, O. Hovorka, C. H. Marrows and H. Fangohr, Appl. Phys. Lett. {\bf 108}, 132406 (2016).

\bibitem{fnote1} DMI suitable for $D_n$ symmetric crystals, ${\cal E}_{{\rm DM}} =-D \left( L_{zx}^{(y)} + L_{yz}^{(x)} \right)$, leads to a Bloch type skyrmionic solution with $\eta =  \pi/2$.




	
\bibitem{Suppli} See supplementary material for this manuscript. 
  
 

\bibitem{Banerjee14} S. Banerjee, J. Rowland,O. Erten and M. Randeria,  Phys. Rev. X {\bf 4}, 031045 (2014). 

\bibitem{Rowland16} J. Rowland, S. Banerjee, and M. Randeria,  Phys. Rev. B {\bf 93}, 020404(R) (2016). 


\bibitem{Herve18} M. Herv\'e, B. Dup\'e, R. Lopes, M. B\"ottcher, M. D. Martins, T. Balashov, L. Gerhard, J. Sinova and W. Wulfhekel, Nature Comm. {\bf 9}, 1015 (2018).
	
\bibitem{Leonov17} A. O. Leonov and I. K\'ezsm\'arki, Phys. Rev. B {\bf 96}, 014423 (2017).


\bibitem{fnote2} A similar anisotropy in DMI corresponding to thin films fabricated with $D_{2d}$ symmetric crystals given by energy density
 	${\cal E}_{{\rm DM}}^- = -D (1+\lambda \cos 2\phi) \left( L_{xz}^{(x)} - L_{yz}^{(y)} \right)$ will produce similar results with reversing roles of Sks and ASks. 

\bibitem{Lobanov16}  I. S. Lobanov, H. J\'onsson and V. M. Uzdin,  Phys. Rev. B {\bf 94}, 174418 (2016).


\vspace{1cm}



{\bf Supplemental Material for ``Theory of skyrmion, meron, anti-skyrmion and anti-meron in chiral magnet"}
	


\setcounter{equation}{9}
\setcounter{figure}{4}

\maketitle


\vspace{0.5cm}

{\bf Euler equation and its skyrmion solution}:\\

Energy for a two dimensional chiral magnet is given by 
\begin{equation}
E = \int d^2\bm{r} \left[ \frac{J}{2} \left( \nabla \hat{m} \right)^2 + {\cal E}_{DM} + H(1-\hat{m}_z) - A(1-\hat{m}_z^2) \right]
\end{equation}
where energy density for DMI, ${\cal E}_{DM} = -D (L_{xz}^{(x)} +L_{yz}^{(y)})$ for $C_{nv}$ symmetric, $-D (L_{yz}^{(x)} +L_{zx}^{(y)})$ for $D_n$ symmetric and $-D (L_{xz}^{(x)} -L_{yz}^{(y)})$ for $D_{2d}$ symmetric systems and we distinguish them by introducing a parameter $\alpha$  with respective values 1, 2 and $-3$. Unit magnetization can be parametrized through spherical variables $(\Theta, \Phi)$ as
$$\hat{m}({\bm{r}}) =\left[ \cos \Phi (\bm{r}) \sin \Theta (\bm{r}),\, \sin \Phi (\bm{r}) \sin \Theta (\bm{r}),\, \cos \Theta (\bm{r}) \right]$$
with $\bm{r}=(r\cos \phi, \, r\sin\phi)$ in polar coordinate. Therefore, the expression of energy reduces to
\begin{eqnarray}
E &=& \int_0^\infty r \, dr \int_0^{2\pi} d\phi \left[ \frac{J}{2} \left(\Theta_r^2 +\frac{\sin^2\Theta}{r^2} \Phi_\phi^2 \right)  \right. \nonumber \\
&+&  H (1-\cos\Theta) - A \sin^2\Theta \nonumber \\
&+&\left. D \left( \Theta_r \pm \frac{\sin (2\Theta)}{2r} \Phi_\phi\right) \sin (\alpha\frac{\pi}{2}-\phi \pm \Phi) \right]
\label{Ener0}
\end{eqnarray} 
where $\Theta_r = \frac{d\Theta(r)}{dr}$ and $\Phi_\phi = \frac{d\Phi(\phi)}{d\phi}$ and
assuming $\Theta (\bm{r}) = \Theta (r)$ and $\Phi (\bm{r}) = \Phi (\phi)$.
In the last term (\ref{Ener0}), positive sign refers to $\alpha = 1$ and $2$ and negative sign refers to $\alpha = -3$. 
Considering $\Phi (\phi) = \zeta \phi + \eta$ with $\zeta = 1$ for $\alpha =1,\, 2$ and $\zeta = -1$ for $\alpha =-3$, and $\eta = 0$ for $\alpha = 1,\, -3$ and $\pi/2$ for $\alpha = 2$, we find Euler equation of $\Theta (r)$ being independent on $\alpha$ as
\begin{eqnarray}
J \left( \Theta_{rr} + \frac{\Theta_r}{r}- \frac{\sin \Theta \cos \Theta}{r^2} \right) + \frac{2D}{r}\sin^2\Theta && \nonumber \\
= H\sin\Theta - A \sin(2\Theta) && 
\label{Diff_eq1}
\end{eqnarray}
where $\Theta_{rr} = \frac{d^2\Theta}{dr^2}$. The corresponding energy is given by 
\begin{eqnarray}
E_{{\rm sk}} &=& 2\pi\int_0^\infty r \, dr  \left[ \frac{J}{2} \left(\Theta_r^2 +\frac{\sin^2\Theta}{r^2}  \right) - A \sin^2\Theta  \right. \nonumber \\
&& + H (1-\cos\Theta) 
+\left. D \left( \Theta_r + \frac{\sin (2\Theta)}{2r} \right) \right]
\label{Ener1}
\end{eqnarray}
Short-distance singularity in exchange energy may be avoided with the boundary condition $\Theta (r=0) = 0$ or $\pi$. We look for skyrmion solution of Eq.~(\ref{Diff_eq1}) by introducing another boundary condition $\Theta (r=\infty) = 0$ (assuming the background as a polarized ferromagnet) along with $\Theta (0) = \pi$. 


In the absence of anisotropy $(A=0)$, an approximate and asymptotically $(r\to 0,\,\infty$) exact analytical solution of Eq.~(1) may be obtained as the exact solution of the simple sine-Gordon like equation $J\frac{d^2\Theta}{dr^2} =   H \sin \Theta $, {\it i.e.},
$\Theta (r) = 4 \arctan \left(\exp \left(-r/r_0\right) \right)$ with 
characteristic length scale $r_0 = \sqrt{J/H}$.
Therefore, an approximate  (exact for $r\to 0$ and $\infty$) solution for $A\neq 0$ may be obtained by considering a reduced form of Eq.~(1) as $J\frac{d^2\Theta}{dr^2} =   H \sin \Theta -A \sin (2\Theta)$ whose solution satisfies an integral equation
\begin{equation}
\int \frac{d\Theta}{\sin (\Theta /2) \sqrt{1-\gamma \cos^2(\Theta /2)}} = -2  (r/r_{_0})
\end{equation}
with $\gamma = 2A/H\, ( \gamma  <1)$, expressible  into
an algebraic equation
\begin{eqnarray}
& &  \sqrt{ 1+ \frac{2 \gamma \cos (\Theta /2)}{1-\gamma \cos (\Theta /2) + \sqrt{1-\gamma}\sqrt{1-\gamma \cos^2(\Theta /2)} }}  \nonumber \\
& & \times \tan (\Theta/4) \hspace{1cm} = \exp \left(-\sqrt{1-\gamma}\,\,r/r_0\right)  \, .
\label{S1}
\end{eqnarray} 
We find the solution of Eq.(\ref{S1}) as 
\begin{equation}
\Theta (r) = 4 \arctan \left(\exp \left(-g(\gamma)\,r/r_0 \right) \right)
\label{S2}
\end{equation}
with $g(\gamma) \simeq 1-\frac{\gamma}{7}- \frac{\gamma^2}{30}$. However, as the smooth change in the orientation of spin depends on $D$, it is natural that we consider another length scale $r_{d} = J/ D $ which is the appropriate length scale for spin-spirals.
By introducing $r_d$ and transforming $r \to r_d \rho$, we find the reduced form of Eq.~(\ref{Diff_eq1}) as
\begin{eqnarray}
\frac{d^2\Theta}{d\rho^2} + \frac{1}{\rho}\frac{d\Theta}{d\rho}- \frac{\sin \Theta \cos \Theta}{\rho^2}  + \frac{2}{\rho}\sin^2\Theta && \nonumber \\
= \frac{H}{H_0}\left( \sin\Theta - \gamma \sin(\Theta)\cos (\Theta) \right)&& 
\label{Diff_eq2}
\end{eqnarray} 
where $H_0 = A_0 = D^2/J$. Numerical solution of Eq.~(\ref{Diff_eq2}) shown in Fig.~\ref{FigS1} for different values of the parameters $H/H_0$ and $\gamma = 2A/H$. We note that while the long-distance solution is independent on these parameters, the short and intermediate distance behavior is strongly parameter dependent, suggesting $r_d$ is not the natural length scale of the system. 


\begin{figure}
	\centering	\includegraphics[width=8.5cm]{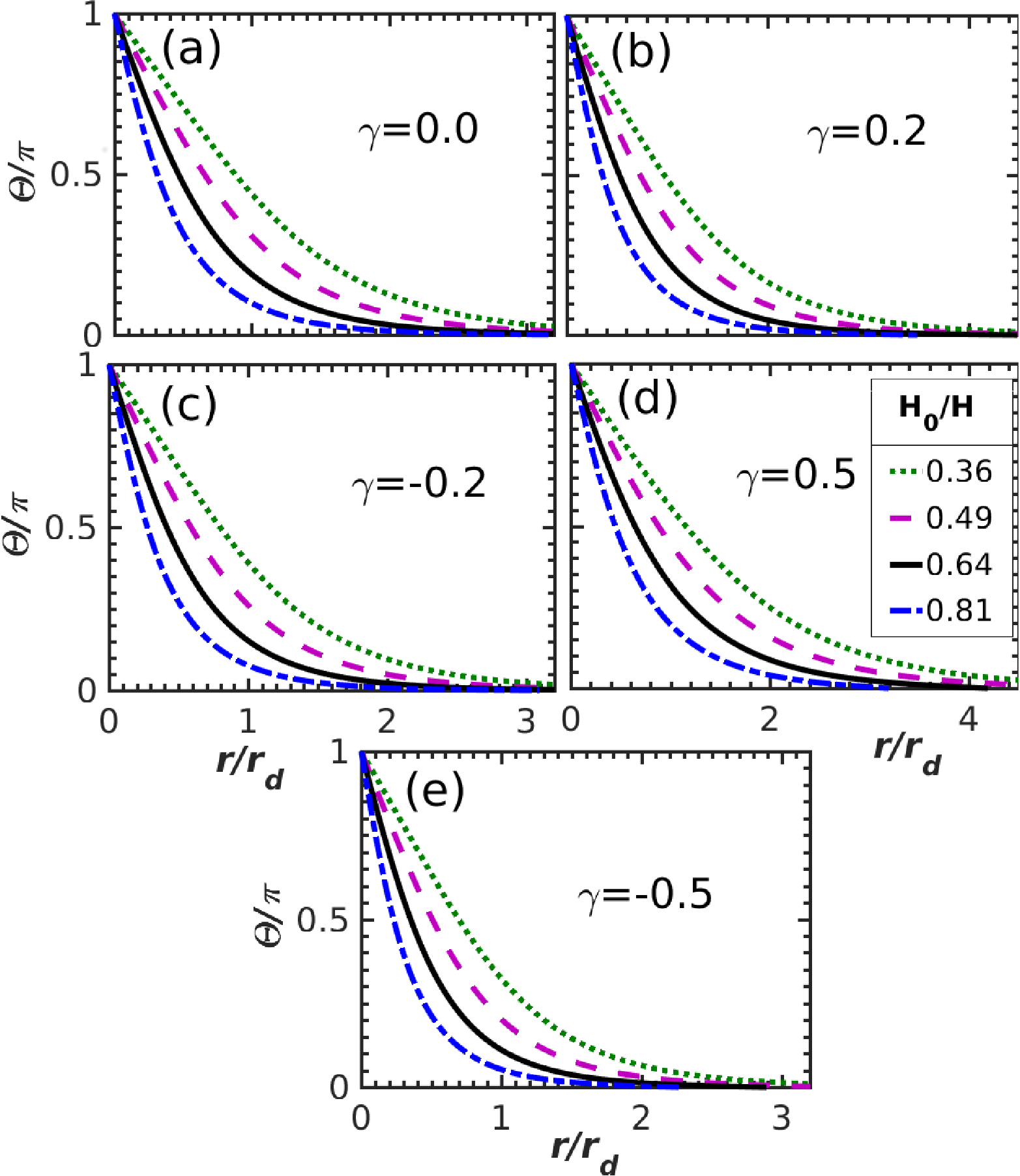}
	\caption{Skyrmion solution: Numerical solution of the Euler equation (\ref{Diff_eq2}), {\it i.e.}, $\Theta (r)$ {\it vs.} $r/r_{d}$ for different values of $\gamma$ in the panels (a)--(e) for same set of $H_0/H$, {\it viz}, 0.36, 0.49, 0.64 and 0.81.    }
	\label{FigS1}
\end{figure} 


We next introduce a length scale $r_s = D /H$ and rescaling $r \to r_s \rho$, we obtain
\begin{eqnarray}
&&\frac{d^2\Theta}{d\rho^2} + \frac{1}{\rho}\frac{d\Theta}{d\rho}- \frac{\sin \Theta \cos \Theta}{\rho^2}   \nonumber \\
&& =  \frac{H_0}{H} \left( -\frac{2}{\rho}\sin^2\Theta 
+\sin\Theta - \gamma \sin\Theta \cos\Theta \right)   
\label{Diff_eq3}
\end{eqnarray} 
whose numerical solution (Fig.~1 of the article and Fig.~\ref{FigS2} below) is almost $H_0/H$ independent for a reasonable range. We thus find natural length scale of the system as $r_s$. Together with the solution of $\Theta (r)$, (i) $\Phi = \phi$ for $\alpha =1$, (ii) $\Phi = \phi +\pi/2$ for $\alpha =2$ and (iii) $\Phi = -\phi$ for $\alpha = -3$ respectively construct magnetic structures of Ne\'el type skyrmion, Bloch type skyrmion and Ne\'el type anti-skyrmion.\\

\begin{figure}
	\centering	\includegraphics[width=8.5cm]{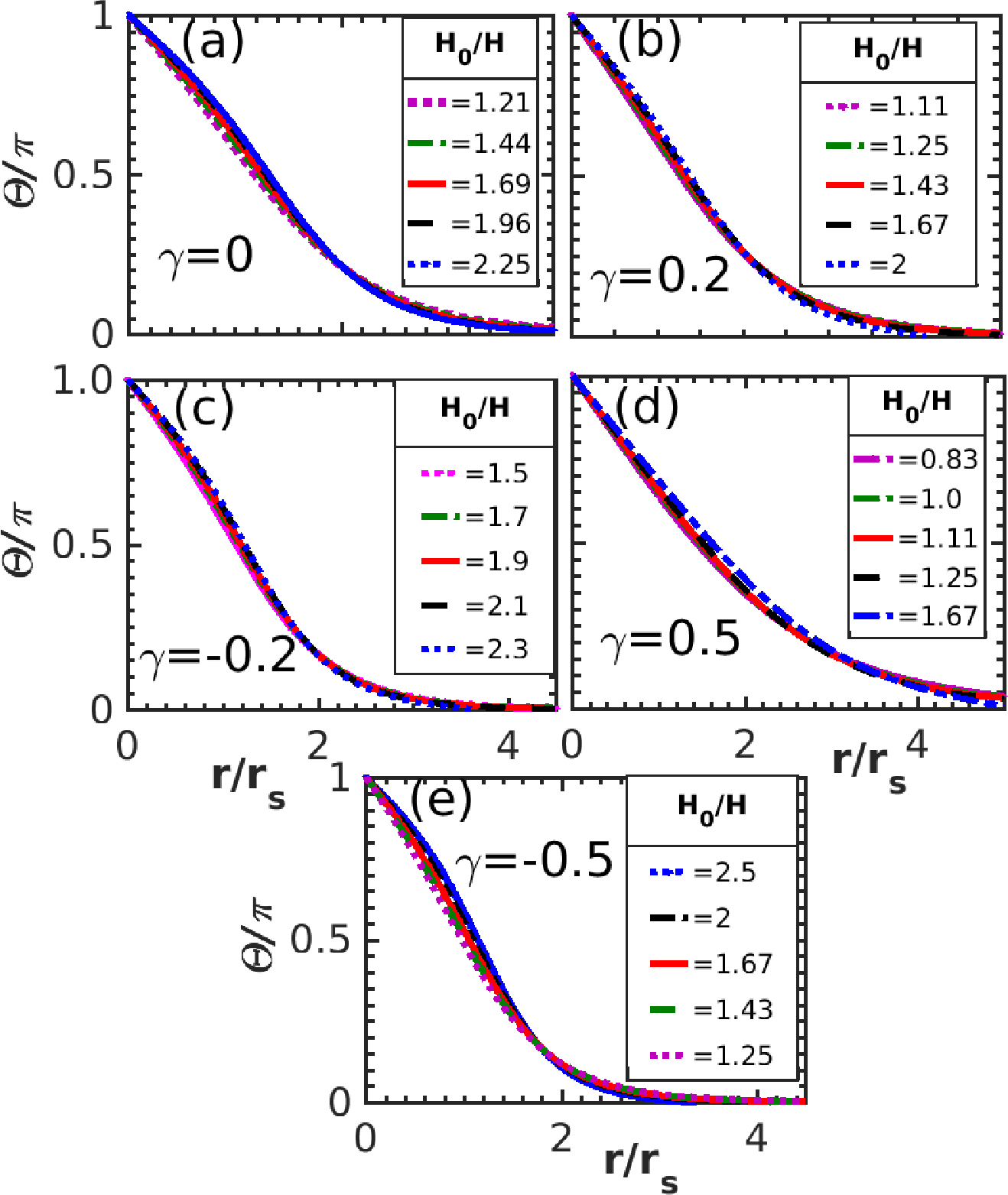}
	\caption{Skyrmion solution: Numerical solution of the Euler equation (\ref{Diff_eq3}), {\it i.e.}, $\Theta (r)$ {\it vs.} $r/r_{s}$ for different values of $\gamma$ and $H_0/H$ in the panels (a)--(e), covering most of the skyrmion phase shown in Fig.~3(a).    }
	\label{FigS2}
\end{figure}

However, the nature of the solution of Eq.(\ref{Diff_eq3}) changes with $A/A_0$ for low field regime. For a moderate to large $H/H_0$, any amount of negative (easy-axis) anisotropy provide normal skyrmion solution (Fig.~\ref{FigS3}(a)) as shown  in Fig.~1 and Fig.~6. When magnetic filed is low and $\vert A\vert /A_0 \lesssim \pi/2\sqrt{2}$ (Dzyaloshinskii criterion \cite{Dzyaloshinskii65} for non-collinear state at zero $H$), the nature of the solution is `chiral-bubble' like \cite{Fert}--where very slow change of $\Theta$ occurs near $r=0$, as shown in Fig.~7(b). The nature of the solution changes for $\vert A\vert /A_0 \gg \pi/2\sqrt{2}$ at low $H/H_0$ from chiral bubble to `meta-stable' skyrmion, as shown in Fig.~7(c), for which $\Theta$ sharply falls near $r=0$. 
The behavior of metastable skyrmions are, however, fundamentally different from the normal skyrmions presented in the main text. While the normal skyrmions are appropriately scaled with $r_s$, the metastable skyrmions are better suited with the length scale $r_d$ (see Fig.~\ref{FigS4}).
Further in contrary to the metastable skyrmions, the radius of a normal skyrmion increases rapidly with the decrease of $H$. The metastable skyrmions are energetically unfavorable to polarized ferromagnet.



\begin{figure}
	\centering	\includegraphics[width=8.5cm]{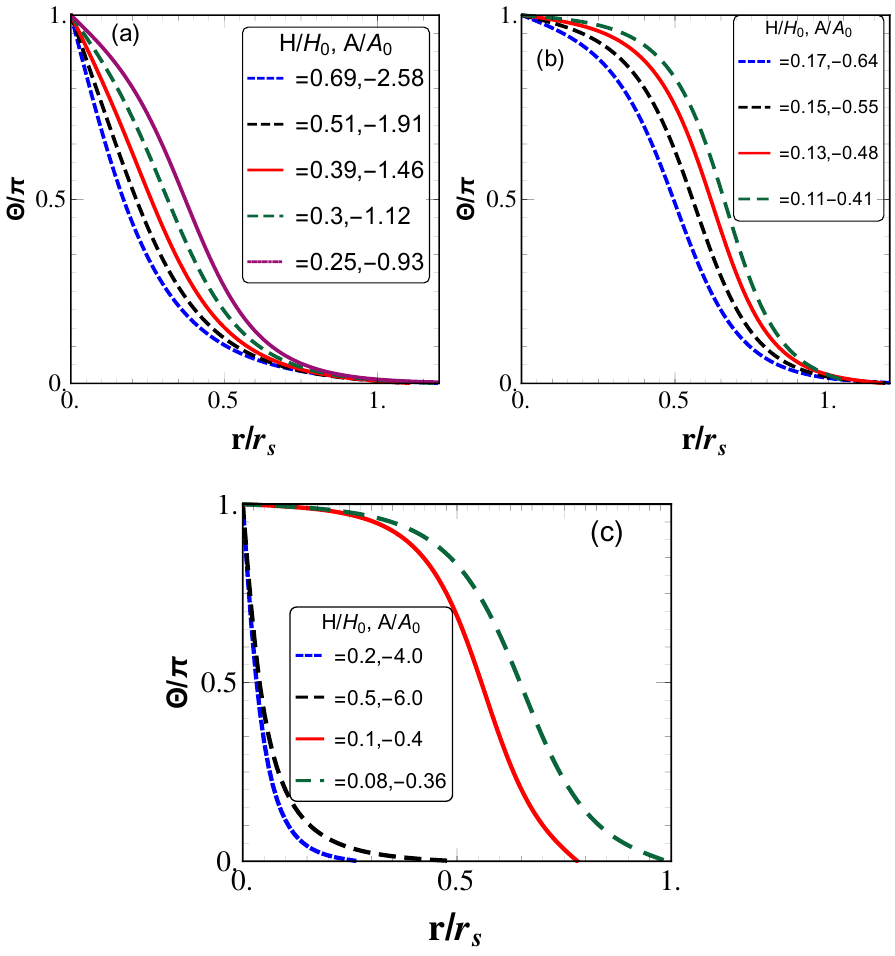}
	\caption{Numerical solution of the Euler equation (\ref{Diff_eq3}), {\it i.e.}, $\Theta (r)$ {\it vs.} $r/r_{s}$ for different values of $H/H_0$ and $A/A_0$. The nature of the solutions changes with the parameters: These are (a) skyrmions (b) chiral bubbles (c) meta-stable skyrmions (also chiral bubbles).    }
	\label{FigS3}
\end{figure}


\begin{figure}
	\centering	\includegraphics[width=8.5cm]{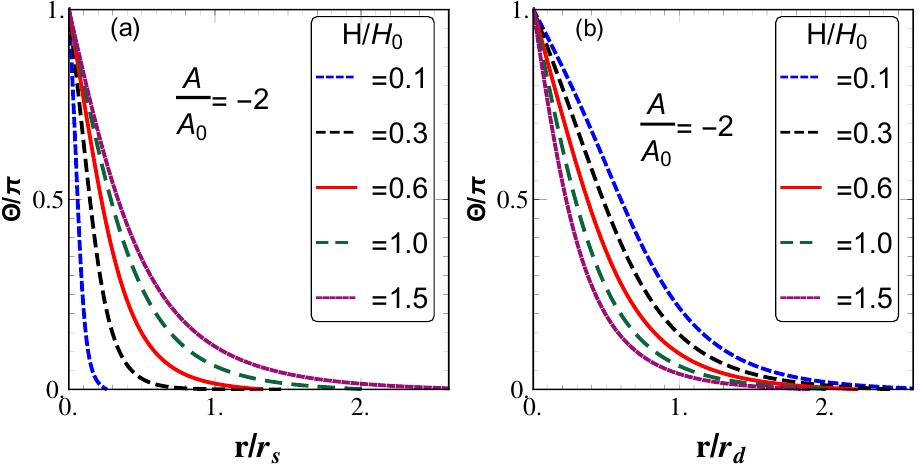}
	\caption{Numerical solution of the Euler equations  (\ref{Diff_eq3}) and (\ref{Diff_eq2}), {\it i.e.}, $\Theta (r)$ {\it vs.} $r/r_{s}$ and $r/r_d$ in 9(a) and (b) respectively for different values of $H/H_0$ and $A/A_0$. Clearly, these metastable skyrmion solutions are better suited with the length scale $r_d$.    }
	\label{FigS4}
\end{figure}



\noindent {\bf Euler equation and its meron solution:}\\


The Euler equation (\ref{Diff_eq1}) will reduce to  
\begin{equation}
J \left( \Theta_{rr} + \frac{\Theta_r}{r}- \frac{\sin \Theta \cos \Theta}{r^2} \right) + \frac{2D}{r}\sin^2\Theta 
+ A \sin(2\Theta) = 0 
\label{Diff_eq4}
\end{equation}
in the absence of magnetic field. For a sufficiently high easy-plane anisotropy $(A>0)$, all the spins will align in the plane (planar ferromagnet). This indicates a boundary condition $\Theta (\infty) = \pi/2$ which together with another boundary condition $\Theta (0)= 0 $ or $\pi$ will provide a solution for meron. By introducing a length scale $r_a =  D  /A$ and rescaling $r \to r_a \rho$, Eq.~(\ref{Diff_eq4}) will reduce to 
\begin{equation}
\frac{d^2\Theta}{d\rho^2} + \frac{1}{\rho}\frac{d\Theta}{d\rho}- \frac{\sin \Theta \cos \Theta}{\rho^2}   
= - \frac{A_0}{A} \left(  \frac{2}{\rho}\sin^2\Theta +\sin(2\Theta) \right)
\label{Diff_meron}
\end{equation}
Taking cue of the skyrmion solution, we assume the solutions of meron is in the form
\begin{equation}
\Theta (r) = \pm \frac{\pi}{2} + 2 \arctan (\exp(-\zeta r/r_a))
\end{equation}
where positive (negative) sign corresponds to spin down (up) at the center of the meron, and the parameter $\zeta$ to be determined by minimizing the corresponding energy. The energy of a meron is given by 
\begin{eqnarray}
E_{{\rm meron}} &=& 2\pi\int_0^\infty r \, dr  \left[ \frac{J}{2} \left(\Theta_r^2 +\frac{\sin^2\Theta}{r^2}  \right) + A \cos^2\Theta  \right. \nonumber \\
&& 
+\left. D \left( \Theta_r + \frac{\sin (2\Theta)}{2r} \right) \right]
\label{Ener_meron}
\end{eqnarray}
which may be simplified to
\begin{equation}
E_{{\rm meron}} = \frac{J}{2}\left[ \ln (2) + I_1\right] + \frac{D^2}{A} \left[ \frac{\ln (2)}{\zeta^2} - \frac{I_2}{\zeta} \right] 
\end{equation}
where $I_1 = \int_0^{\infty} (1/r) \tanh^2(r) dr$, $I_2 = \int_0^\infty \left( \frac{\tanh r}{\cosh r} + \frac{r}{\cosh r}\right) dr = 1+2G$, and Catalan's constant $G = \sum_{k=0}^\infty \frac{(-1)^k}{(2k+1)^2}\approx 0.91$.








\begin{thebibliography}{99}
	\bibitem[S1]{Dzyaloshinskii65} I. E. Dzyaloshinskii, Sov. Phys. JETP {\bf 20}, 665 (1965).	
	\bibitem[S2]{Fert} See, for example, A. Fert, N. Reyren, and V. Cros, Nature Review Materials {\bf 2}, 17031 (2017).
	
\end{thebibliography}
\end{document}